\documentclass[useAMS,usenatbib,usegraphicx,letterpaper]{mn2e}
\title[Multi-band polarimetry of the 3C~15 jet]{The structure of the jet in 3C~15 from multi-band polarimetry}
\author[F. Dulwich et al.]{F.~Dulwich$^{1}$, D.M.~Worrall$^{1}$, M.~Birkinshaw$^{1}$, C.A.~Padgett$^{2}$, E.S.~Perlman$^{2}$ \\
    $^{1}$ H.H. Wills Physics Laboratory, University of Bristol, Tyndall Avenue, Bristol BS8 1TL, UK \\
    $^{2}$ Joint Center for Astrophysics, University of Maryland-Baltimore County, 1000 Hilltop Circle, Baltimore, MD 21250, USA}

\begin{document}

\date{Received; Accepted.}

\pagerange{\pageref{firstpage}--\pageref{lastpage}} \pubyear{2006}

\maketitle

\label{firstpage}

\begin{abstract}
We investigate the structure of the kpc-scale jet in the nearby ($z=0.073$) radio galaxy 3C~15, using new optical {\it Hubble Space Telescope} ({\it HST}) ACS/F606W polarimetry together with archival multi-band {\it HST} imaging, {\it Chandra} X-ray data and 8.4~GHz {\it VLA} radio polarimetry. The new data confirm that synchrotron radiation dominates in the optical. With matched beams, the jet is generally narrower in the optical than in the radio, suggesting a stratified flow. We examine a simple two-component model comprising a highly relativistic spine and lower-velocity sheath. This configuration is broadly consistent with polarization angle differences seen in the optical and radio data. The base of the jet is relatively brighter in the ultraviolet and X-ray than at lower energies, and the radio and optical polarization angles vary significantly as the jet brightens downstream. Further out, the X-ray intensity rises again and the apparent magnetic field becomes simpler, indicating a strong shock. Modelling the synchrotron spectrum of this brightest X-ray knot provides an estimate of its minimum internal pressure, and a comparison with the thermal pressure from X-ray emitting gas shows that the knot is overpressured and likely to be a temporary, expanding feature.

\end{abstract}


\section{Introduction}\label{sec:intro}
The jets and lobes of radio galaxies are sites of some of the most energetic particles in the Universe. Many nearby sources show complex structures at all angular sizes, and most have been observed to emit synchrotron radiation from X-ray to radio wavelengths in their resolved jets. Much of the physics of these jets is unclear, although it is generally believed that they are relativistic, collimated outflows of plasma, launched from an active galactic nucleus by some largely unknown mechanism. The plasma contains magnetic fields that play significant roles in shaping the jets and producing the radiation we detect.

Early radio observations of kpc-scale jets revealed that the emissions were of synchrotron origin and often highly polarized, implying ordered magnetic fields within the jet \citep[e.g.][]{BP84}. More recent observations have revealed optical counterparts to some radio jets, often showing a close correspondence in radio and optical morphology. Combined with optical spectra, this suggested that the optical emission was also synchrotron in nature \citep[e.g.][and references therein]{Keel88,FC96}. The kpc-scale jets of most nearby radio galaxies observed with the {\it Chandra X-ray Observatory} ({\it CXO}) are detected in X-rays, and it has been argued that the dominant emission process is synchrotron radiation \citep*{Worrall01,Hardcastle01}. Since the lifetime of high-energy electrons to synchrotron-radiation losses is short (tens to hundreds of years for electrons with $\gamma = E/mc^2 \sim 10^7 - 10^8$ in a magnetic field of a few nT), it is clear that there must be significant localised particle acceleration occurring throughout the jet to maintain the observed emissions.

Magnetic-field structures are indicated primarily by the polarization pattern of the synchrotron radiation: polarization vectors show the direction of the apparent magnetic field, while the fraction of polarized emission indicates the ordering of the field. Multi-band imaging and polarimetry therefore tell us about the magnetic fields seen by emitting particles of different energies.

\begin{figure}
\centering
\includegraphics[width=84mm]{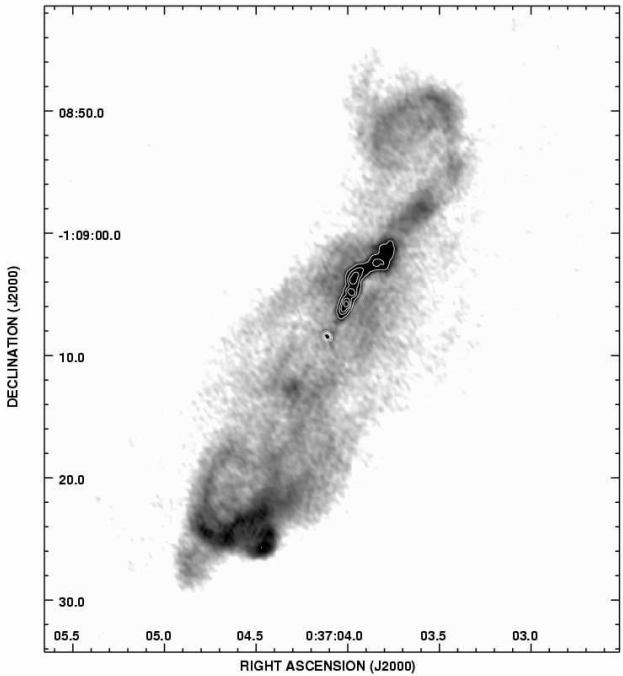}\\
\caption{8.4~GHz image of 3C~15 made from archival A- and C-configuration {\it VLA} data, showing the core, NW jet, and lobes.}
\label{fig:radio}
\end{figure}

3C~15 is a bright, compact and relatively nearby radio galaxy at redshift $z=0.073$, and it has an elliptical host galaxy showing weak optical emission lines \citep{Tadhunter93}. At 8.4~GHz it is dominated by a luminous, one-sided jet, although it also exhibits diffuse emission from extended radio lobes and a weak counter-jet \citep[][and Figure~\ref{fig:radio}]{Leahy97}. The bright core is not resolved with the {\it VLA}. The radio morphology of 3C~15 is unusual: the northern jet is knotty and appears similar to those in Fanaroff--Riley (FR) class~I sources, while the southern jet is barely detected in the radio. However, it powers a `warm spot' in the southern radio lobe which is typical of a more powerful FR~class~II galaxy. The radio power of 3C~15 \citep[$L_{\textrm{\tiny 178 MHz}} = 3.3 \times 10^{25}$ W Hz$^{-1}$ sr$^{-1}$,][]{Hardcastle98} is close to the FR~I/FR~II boundary at $10^{25}$ W Hz$^{-1}$ sr$^{-1}$ \citep{Ledlow92}.

An optical jet in 3C~15, closely following the morphology of the northern radio jet, was detected with the {\it Hubble Space Telescope}, and it has been suggested that synchrotron radiation dominates in both the radio and optical wavebands \citep{Martel98}. The radio--optical spectral index of the integrated jet emission $\alpha_{ro}=0.95 \pm 0.01$ was the softest known at that time. The jet and lobes are detected in X-rays with the sub-arcsecond resolution offered by {\it Chandra}. \citet{Kataoka03} presented four different models of the spectral energy distribution (SED) at the brightest X-ray knot in the jet; they favour a broad-band synchrotron spectrum with an equipartition magnetic field, and a cooling break at a Lorentz factor $\gamma_{brk} = 2.6 \times 10^4$. We agree with the general characteristics of this model, although the updated calibration products released by the Chandra X-ray Center (CXC) allow us to extract a more reliable X-ray spectrum of this region of the jet, and hence model the spectrum of the knot with greater confidence. Multi-band optical photometry also allows us to place a good constraint on the optical spectral index $\alpha_o$ in the knot.

In this paper we re-examine the 3C~15 jet, using new {\it HST} optical polarimetry in conjunction with the reprocessed radio and X-ray data. Section~\ref{sec:observations} describes the multi-waveband observations we used. We present our results in Section~\ref{sec:results}, and in Section~\ref{sec:discussion} we examine simple models of the jet and its flow. Detailed magnetic modelling of the jet is deferred to a later paper. Our primary aim is to use the apparent magnetic-field structure to infer characteristics of the plasma flow and locate sites of particle acceleration. Throughout, we adopt values for the cosmological parameters $H_0=70$ km s$^{-1}$ Mpc$^{-1}$, $\Omega_{{\Lambda}0}=0.7$ and $\Omega_{m0}=0.3$. At the redshift of 3C~15, one arcsecond corresponds to a projected distance of 1.39 kpc.

\section{Observations}\label{sec:observations}
\subsection{Radio data}\label{subsec:radiodata}
Archival X-band (8.4~GHz) {\it VLA} data of 3C~15 were reprocessed using the {\sc aips} package from NRAO. We used data from a 20 minute observation in the C-configuration on 1989 September 14, and those from an additional 87 minute A-configuration (high-resolution) observation on 1990 May 26.

The data were read into {\sc aips} and bad scans were flagged and removed. Antennas 26 and 29 were not present for the observation made using the A-array, while antennas 15 and 23 were removed due to high noise across all baselines. In the C-configuration, antennas 11, 23 and 29 were absent and antennas 6, 18 and 22 required further flagging. The first ten seconds of all scans were also removed to avoid problems with bad visibilities. Flux-density calibration used 3C~286 for the A-configuration data and 3C~48 for the C-configuration data, while both observations made use of 0040-017 for phase referencing.

\begin{figure*}
\begin{minipage}{180mm}
\centering
\includegraphics[width=180mm]{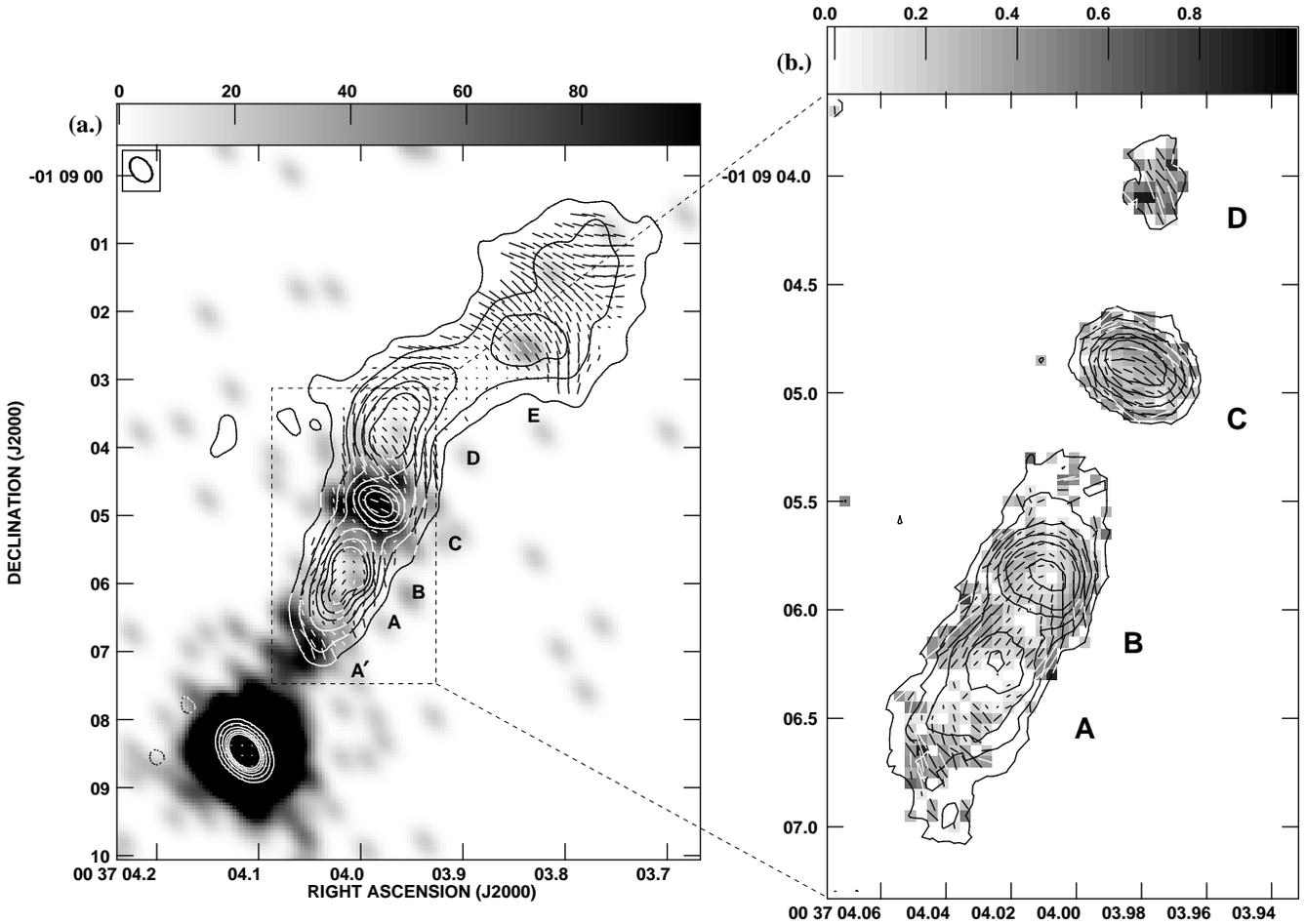}\\
\caption{Left (a.): $0.40 \times 0.27$ arcsec resolution X-band (8.4 GHz) radio contours, and percentage polarization vectors rotated by 90$\degr$ to indicate the apparent magnetic field in the radio jet. Contour levels are $0.2 \times (-1, 1, 4, 10, 18, 26, 36)$ mJy beam$^{-1}$, and a vector of length 0.5 arcsec corresponds to 100 per cent polarization. Negative contours are dashed. The jet and nucleus are also clearly seen in the X-ray image of the full \textit{Chandra} energy band, resampled by a factor of 4 and smoothed with the same $0.40 \times 0.27$ arcsec Gaussian (shown in the heavily-clipped grey scale image). Note, however, that the point-spread function for the unconvolved X-ray image has an FWHM of 0.65 arcsec. Right (b.): High-resolution optical polarimetry from {\it HST}, where a vector of length 0.2 arcsec corresponds to 100 per cent polarization and the grey-scale indicates the fraction. Contours show the optical intensity, with levels $0.05 \times (1, 1.8, 2.8, 4.2, 5.8, 11, 14.1)$ electrons sec$^{-1}$. Polarization vectors are rotated by 90$\degr$ to indicate the apparent magnetic field direction. The ACS/F606W resolution is diffraction limited at 0.06 arcsec.}
\label{fig:data}
\end{minipage}
\end{figure*}

After determining the intrinsic antenna polarization using {\sc pcal} with the calibrator sources, both data sets were phase and amplitude self-calibrated to good convergence using {\sc imagr} and {\sc calib} in {\sc aips}. The A- and C-array data were then combined using the method described by \citet{Black92}, and a further self-calibration was performed on the new database. The large-scale intensity map was deconvolved using the maximum entropy method (MEM) embodied in the {\sc aips} task {\sc vtess}, since it handles extended structure well. However, as pointed out by \citet{Black92}, MEM algorithms do not give good results near bright, compact regions. For the extended structure, we employ their hybrid technique of an initial shallow Clean, followed by 50 iterations of {\sc vtess} applied only to the residuals. The convolved Clean components were then restored to the {\sc vtess} image to give the Stokes $I$ map (Figure~\ref{fig:radio}).

The radio emission from the jet was also imaged at the highest angular resolution in Stokes $I$, $Q$ and $U$ using A-configuration data only. A straightforward Clean/self-calibration cycle was used without the aid of {\sc vtess} in this case.

All the radio images were made on an 0.05 arcsec grid, and reconstructed using a synthesised elliptical Gaussian beam of 0.40 $\times$ 0.27 arcsec with major axis in position angle 35$\degr$. This gives a noise level in Stokes $I$ of $\sim 40$~{$\mu$}Jy beam$^{-1}$ from both the A- and C-arrays. The percentage-polarization and apparent Magnetic Field Position Angle (MFPA) images were computed using $100 \times (Q^2+U^2)^{\frac{1}{2}}/I$ and $\frac{1}{2}\tan^{-1}(U/Q)+90\degr$, respectively. The Rician bias in the percentage polarization \citep{Serkowski62} was corrected by measuring the noise levels in the Stokes $Q$ and $U$ images and using the {\sc polc} algorithm in the {\sc aips} task {\sc comb}. Component fluxes were calculated by fitting two-dimensional Gaussians using the task {\sc imfit} in {\sc aips}.

\subsection{Optical data}\label{subsec:opticaldata}
Optical {\it HST} images of 3C~15 were obtained on 2003 December 9 with the ACS/WFC instrument using the wide-band F606W filter and the three optical polarizers, each in a 2872 second exposure. Data were flat-fielded and bias-corrected using standard packages in {\sc iraf/stsdas}, and {\sc crreject} was run to remove cosmic-ray events \citep[see][for further details]{Perlman06,Perlman99}. The ACS images were combined using multidrizzle \citep{Koekemoer02}, allowing for corrections in the chip geometry \citep{AK04}. Stokes $I$, $Q$ and $U$ images were produced and combined following the guidelines in the ACS Data Handbook \citep{Pavlovsky05}, and convolved with a $\sigma=1$ pixel circular Gaussian. The optical percentage-polarization and apparent MFPA images were produced as described in section~\ref{subsec:radiodata}. Corrections were needed for the optical MFPA to include the {\it HST} roll-angle ($244.9\degr$ for program 9847) and the offset given by the camera geometry \citep[$-38.2\degr$ for WFC, from][]{Pavlovsky05}. The Rician bias in the percentage polarization \citep{Serkowski62} was corrected using a script adapted from the Space Telescope European Coordinating Facility (ST-ECF) package \citep{Hook00} in {\sc iraf}, following \citet{WK74}. The script uses a `most-probable value' estimator, excluding pixels where the signal-to-noise ratio is $<0.5$, or where the most probable value in the percentage polarization is negative or $>100$ per cent. Small corrections were made for the parallel and perpendicular transmittance of each filter.

\begin{table*}
\begin{minipage}{180mm}
\caption{Observation details and flux densities at knot~C.}
\label{tab:fluxes}
\begin{tabular}{llllll}
\hline
Instrument, Band/Filter & ID & Integration Time (s) & Date & Frequency $\nu$ (Hz) & Knot~C flux $S_{\nu}$ (Jy) \\
\hline
{\it VLA} A-config, X-band      & AB534 & $5.2 \times 10^3$  & 1990 May 26 & $8.35 \times 10^9$     & $(22.2 \pm 0.2) \times 10^{-3}$      \\
{\it HST} STIS/CCD, F28X50LP    & 8233  & $4.0 \times 10^2$  & 2001 Jan 27 & $3.97 \times 10^{14}$  & $(2.51 \pm 0.05) \times 10^{-6}$     \\
{\it HST} WFPC2/PC, F702W       & 5476  & $2.8 \times 10^2$  & 1995 Jun 18 & $4.29 \times 10^{14}$  & $(2.2 \pm 0.2) \times 10^{-6}$       \\
{\it HST} ACS/WFC1, F606W + POL & 9847  & $8.6 \times 10^3$  & 2003 Dec 9  & $4.97 \times 10^{14}$  & $(1.87 \pm 0.01) \times 10^{-6}$     \\
{\it HST} WFPC2/PC, F555W       & 6348  & $6.0 \times 10^2$  & 1997 Feb 5  & $5.43 \times 10^{14}$  & $(1.6 \pm 0.1) \times 10^{-6}$       \\
{\it HST} STIS/NUVMAMA, F25QTZ  & 8233  & $7.2 \times 10^3$  & 2001 Jan 27 & $1.21 \times 10^{15}$  & $(6.3 \pm 0.1) \times 10^{-7}$       \\
{\it Chandra} ACIS, 0.3--6 keV  & 2178  & $2.6 \times 10^4$  & 2000 Nov 6  & $2.41 \times 10^{17}$  & $(1.5^{+0.4}_{-0.3}) \times 10^{-9}$ \\
\hline
\end{tabular}
\end{minipage}
\end{table*}

3C~15 was also observed without the polarimeters at various times using a number of other filters and instruments aboard {\it HST}, listed in Table~\ref{tab:fluxes}. Galaxy subtraction was performed using the tasks {\sc ellipse} and {\sc bmodel} in {\sc iraf}, and {\sc imcalc} in {\sc stsdas}. The optical images were subsequently read into {\sc aips} for analysis.

\subsection{X-ray data}\label{subsec:xraydata}
The ACIS-S instrument on {\it Chandra} was used to observe 3C~15 on 2000 November 6 (ObsID~2178) in {\sc faint} imaging mode, using chip S3 with a 1.7-second frame time. The level-1 X-ray data were reprocessed using {\sc ciao}~3.2.1 and {\sc caldb}~3.0.0 from the Chandra X-ray Center to take advantage of recent calibration products. New bad-pixel files were created following the analysis `threads' (online at http://cxc.harvard.edu/ciao/threads/), and a new level-1 event file was created with pixel-randomisation removed. Filtering for good grades (with values 0, 2, 3, 4 and 6) and excluding the interval of a short background flare resulted in a level-2 event file with a total good-time of 25.95 ks.

The X-ray image was resampled by a factor 4, to give a 0.123 arcsec pixel scale. This oversamples the point-spread function of {\it Chandra}, which is $\simeq0.65$ arcsec for our observation. It should be noted that this procedure does not increase the image resolution in any way, although performing a Gaussian smooth ($0.40 \times 0.27$ arcsec) on the resampled image highlights features of the X-ray jet that the eye could otherwise miss (Figure~\ref{fig:data}a, grey-scale).

\begin{figure*}
\begin{minipage}{130mm}
\centering
\medskip
\medskip
\medskip
\includegraphics[width=120mm]{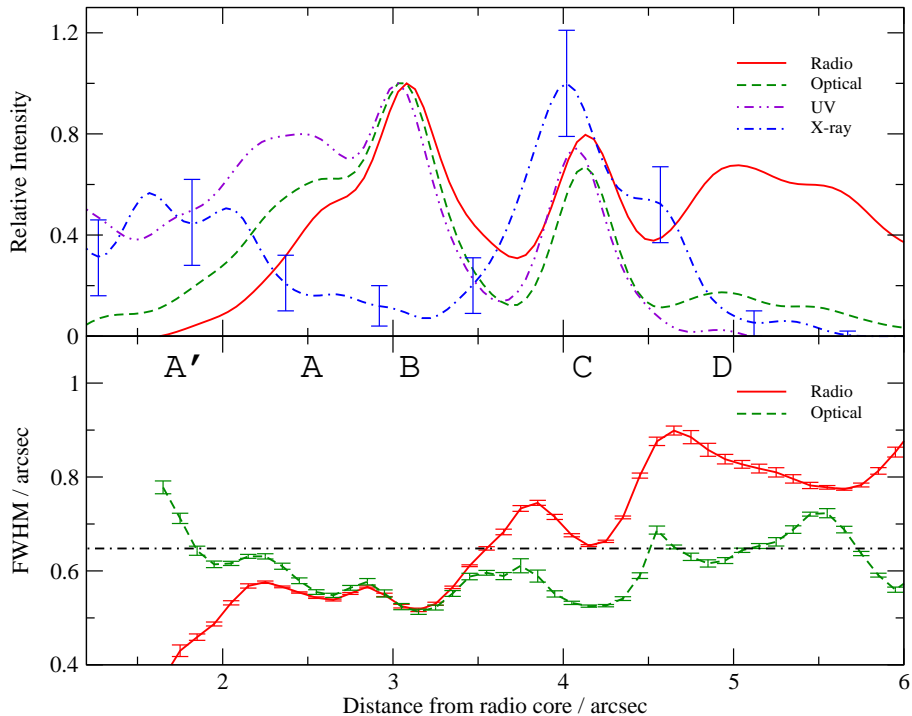}\\
\caption{Top Panel: Linear intensity profiles 2 arcsec wide along the major axis of the inner jet, normalised to unit peak intensity. The profiles are shown for the X-band radio (solid line), V-band optical (dashed; {\it HST} ACS/F606W), the STIS ultraviolet band (dot-dot-dash; {\it HST} NUVMAMA/F25QTZ) and the full {\it Chandra} X-ray band (dot-dash). The images were smoothed with a $0.40 \times 0.27$ arcsec Gaussian, and representative X-ray error bars are shown. Bottom Panel: The full-width half-maximum of one-dimensional Gaussian slices fitted every 0.1 arcsec to the radio and optical data in the direction perpendicular to the jet show that the jet is generally narrower in the optical than in the radio. The dot-dash horizontal line on the bottom panel shows the FWHM of the {\it Chandra} PSF.}
\label{fig:profile}
\end{minipage}
\end{figure*}

\section{Results}\label{sec:results}
Since the absolute astrometry of radio images from the {\it VLA} is more accurate than either the optical or X-ray data, all the cores were registered to the coordinates of the radio peak (at $\rmn{RA}= 00^{\rmn{h}} 37^{\rmn{m}} 04\fs114$, $\rmn{Dec.}= -01\degr 09\arcmin 08\farcs46$ in J2000) prior to any analysis. The optical image was shifted by $(-1.14, -0.44)$ arcsec and the X-ray by $(0, -0.22)$ arcsec in ($\alpha$, $\delta$). Figure~\ref{fig:data}a shows the radio contours overlaid on the corrected X-ray image.

\subsection{Multi-wavelength morphology}
The high-resolution optical image of the jet shows four knots, labelled A--D in Figure~\ref{fig:data}b. We adopt the nomenclature of \citet{Martel98} -- note that \citet{Leahy97} label the radio knots differently. Radio knot~A is resolved into two optical knots, A~and~B, in the {\it HST} image, while knot~D is detected here in the optical for the first time.

Multi-band intensity profiles of the jet out to knot~D were created from slices 2 arcsec wide, after rotating all the images 61.5 degrees west from north to align the inner jet with the $x$-axis. Galaxy subtraction was performed on the optical image (Section~\ref{subsec:opticaldata}), and the ultraviolet intensity profile was created by first subtracting a mean background value from the UV image. The optical and UV images were then convolved with a $0.40 \times 0.27$ arcsec Gaussian to match the radio beam size and orientation. As shown in Figure~\ref{fig:profile}, the radio and optical profiles follow each other closely -- the $\sim20$ mas ($\sim28$ pc) offset between the radio and optical peaks at knots~B and~C is not significant, since this is only $\sim10$ per cent of the radio beam width. However, there is a somewhat larger 0.10 arcsec (140 pc) offset between the broad-band X-ray and radio intensity maxima in the region of knot~C. With 69 X-ray counts in this region, the peak should be located to better than 60~mas with {\it Chandra}; this offset is therefore only a $1.7\sigma$ detection. Thus, we regard all the peak locations of knots~A--D as consistent in the radio, optical, UV and X-ray images.

The bottom panel of Figure~\ref{fig:profile} shows width profiles transverse to the jet direction, created by fitting one-dimensional Gaussian slices every 0.1 arcsec to the radio and optical data. Measurements were taken directly from the radio image (i.e. it was not deconvolved), although as before the optical image was smoothed with a two-dimensional Gaussian to match the radio beam. The profile shows the full-width-half-maximum of each slice, indicating that the jet is generally narrower in the optical than in the radio, except in the region around knots~A and~B where the widths are similar. From both panels, it is apparent that the jet becomes slightly narrower at the flux maxima. The width profiles do not extend inside 2 arcsec, as the galaxy-subtraction procedure used for the optical data introduces artefacts close to the core. A width profile was not generated from the X-ray data, since the size of the {\it Chandra} PSF (indicated on Figure~\ref{fig:profile}, bottom panel) is too large for a fair comparison at this resolution.

The first peak in the X-ray emission at the base of the jet (knot~A$'$, first noted by \citet{Kataoka03} and shown here in Figures~\ref{fig:data}a and~\ref{fig:profile}) occurs closer to the nucleus than any other feature. This knot is located in a region where the optical and radio intensity are still weak, before peaking further downstream at knots~A and~B. The ultraviolet {\it HST} image also shows that the ultraviolet intensity at A$'$ rises faster than in the optical, which in turn rises faster than in the radio. This observation of a kpc-scale jet appearing to turn on in the X-ray before its first intensity flare at lower frequencies has been seen in other FR~I sources \citep[][for 3C~31 and NGC~315 respectively]{Hardcastle02,Worrall06b}.

\begin{figure*}
\begin{minipage}{120mm}
\centering
\includegraphics[width=120mm]{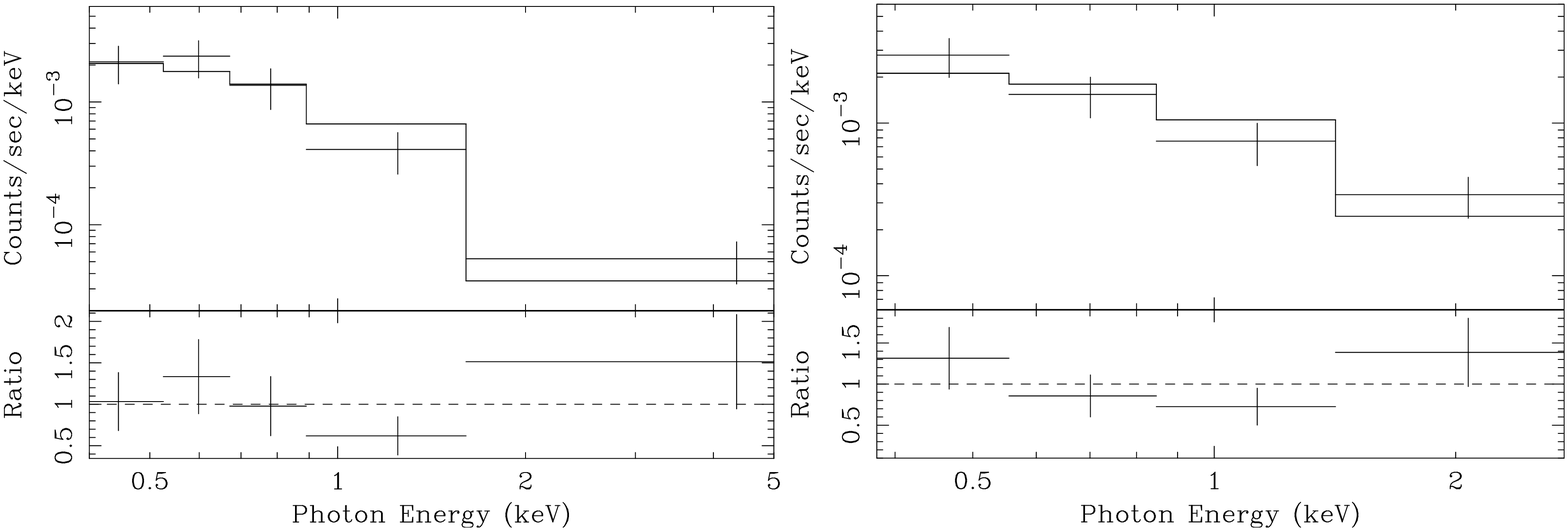}\\
\caption{X-ray spectra of jet knots A$'$ and C, both fitted by a simple power-law absorbed by a Galactic $N_H = 3.03 \times 10^{20}$~cm$^{-2}$. Fitting parameters are given in Table~\ref{tab:XrayFit}. Left (a.): Spectrum of knot A$'$ with 8 counts per bin. Right (b.): Spectrum of knot~C with 12 counts per bin. Note that the energy scales are different and the two spectral slopes are consistent with one another.}
\label{fig:spectra}
\end{minipage}
\end{figure*}

Knot A$'$ provides $\sim40$ X-ray counts, allowing us to extract a low-resolution spectrum, although the photon statistics do not permit fitting a model more complex than a simple power-law. The fitting was performed using the C-statistic in {\sc xspec}. The spectrum, shown in Figure~\ref{fig:spectra}, is well-fitted by a simple power-law with Galactic absorption ($N_H~=~3.03~\times~10^{20}$~cm$^{-2}$, obtained using the {\sc colden} program from the CXC) and photon index $\Gamma_{A'} = 2.7 \pm 0.5$. Despite there being no significant radio or optical emission from A$'$, it is interesting that its X-ray spectrum is consistent with that from knot~C which is bright at all wavelengths and where we argue below that particle acceleration is significant. The steep X-ray spectrum of knot~A$'$ supports the emission being synchrotron from high-energy electrons with short lifetimes, rather than from inverse-Compton scattering of CMB photons. The emission here may be coming from a region whose self-absorption frequency is above 8.4~GHz, which could explain the lack of observed radio emission from A$'$. If the radio-to-X-ray spectrum is the same at A$'$ and C, then knot~A$'$ would be detected clearly in the radio. Since A$'$ is not visible in the radio image, the spectrum of A$'$ cannot be the same as that of knot~C over the entire radio-to-X-ray band.

\begin{table*}
\begin{minipage}{120mm}
\caption{Best-fitting parameters for the X-ray spectra of knot~A$'$ and knot~C in 3C~15.}
\label{tab:XrayFit}
\begin{tabular}{lccccc}
\hline
Knot & Photon Index$^\mathrm{a}$ & 1~keV Normalisation$^\mathrm{b}$ & C-statistic & PHA bins & Fit$^\mathrm{c}$ \\
\hline
A$'$ & $2.7 \pm 0.5$ & $(1.7^{+0.5}_{-0.4}) \times 10^{-6}$ & 3.7 & 5 & 39.4\% \\
C    & $2.4 \pm 0.5$ & $(2.2^{+0.6}_{-0.5}) \times 10^{-6}$ & 3.4 & 4 & 49.8\% \\
\hline
\end{tabular}
\\
$^\mathrm{a}$ Errors are 90 per cent for one interesting parameter.\\
$^\mathrm{b}$ Units are photons~cm$^{-2}$~s$^{-1}$~keV$^{-1}$. Errors are 1$\sigma$ for two interesting parameters.\\
$^\mathrm{c}$ Goodness of fit, estimated using Monte-Carlo methods. A good fit is indicated by values near 50 per cent, while very low percentages suggest the data are over-fitted.
\medskip
\end{minipage}
\end{table*}

The region around knots~A and~B is the only place where the width of the jet is similar in the optical and radio images. The X-ray intensity falls to a minimum in the A/B~region, where the optical and radio emissions show large peaks. However, knot~C is clearly detected in all three wavebands, although it is unresolved in the {\it Chandra} X-ray image. The X-ray spectrum of knot~C was fitted with a broken power law by \citet{Kataoka03}, who found a photon index of 3.0 below $0.90 \pm 0.18$~keV and flattening to 1.5 at higher energies. An alternative fit as a single power law had Galactic absorption and a photon index $\Gamma_C = 1.7 \pm 0.4$, although in this case the reduced $\chi^2$ was 3.1 for two degrees of freedom. It is difficult to justify a complex model when fitting only $\sim50$ counts from knot~C (radius 1.3 arcsec), and our recalibrated data fit a simple power law with Galactic absorption and a photon index $\Gamma_C = 2.4 \pm 0.5$. Background subtraction used an annulus (radii 3.1 and 5.9 arcsec) around the nucleus, thus ensuring any constant surface brightness X-rays from the lobes were subtracted. The fitted spectrum, again using the C-statistic, is shown in Figure~\ref{fig:spectra}, and we note a hint of the same `concave' feature in the residuals at knot~C mentioned by \citet{Kataoka03}. Table~\ref{tab:XrayFit} shows the parameters of our best fit, with a C-statistic of 3.4 using 4 PHA bins. The spectrum shows a low-significance rise at low energies, due to a relatively large number of counts (11) between 520--580~eV.

The size of the {\it Chandra} PSF does not allow us to add the X-ray width profiles to the bottom panel of Figure~\ref{fig:profile}. To compare the width of the jet at knot~C we convolved the optical and radio images to match the {\it Chandra} resolution and fitted Gaussian profiles to slices across the knot. The full-width half-maximum at knot~C was then found to be $0.59\pm0.05$ arcsec, $0.74\pm0.02$ arcsec and $0.89\pm0.01$ arcsec in the matching X-ray, optical and radio images, respectively. The width of the core on these images is the same. The results suggest that knot~C becomes increasingly broader from X-ray to optical to radio wavelengths.

The diffuse X-ray emission from the lobes \citep{Kataoka03} makes it difficult to determine the significance of the faint X-rays that we appear to detect from the outer jet. Knot~D exhibits just one count, although there are three X-ray counts from knot~E (Figure~\ref{fig:data}a). With an expected background of 0.31 counts over the whole {\it Chandra} band at knot~E, the Poisson probability of obtaining three counts where 0.31 are expected is 0.36 per cent, implying a significant detection. Two counts have energies less than 1~keV, although the third is above 8.5~keV and so is likely to be part of the particle background. Repeating the analysis for the same two counts on a 0.3--6~keV image gives a lower background of 0.13 counts, but with a higher chance detection of 0.74 per cent.

Figure~\ref{fig:profile} shows the jet staying of constant width or narrowing slightly in the optical as it widens in the radio. The upstream edges of knots~A and~B are slightly brighter in the optical than in the radio, while the situation is reversed at the downstream edges -- similar behaviour has been observed by \citet{Perlman99} in knots in the M~87 jet. Downstream of the X-ray peak at knot~C, the jet exhibits a kink of approximately 20$\degr$ in both the optical and radio maps. Although the radio jet is seen to extend further and change direction again (twice), the optical emission fades rapidly after knot~C, and little optical emission is detected after knot~D.

\subsection{Jet polarimetry}
\begin{figure*}
\begin{minipage}{160mm}
\centering
\includegraphics[width=160mm]{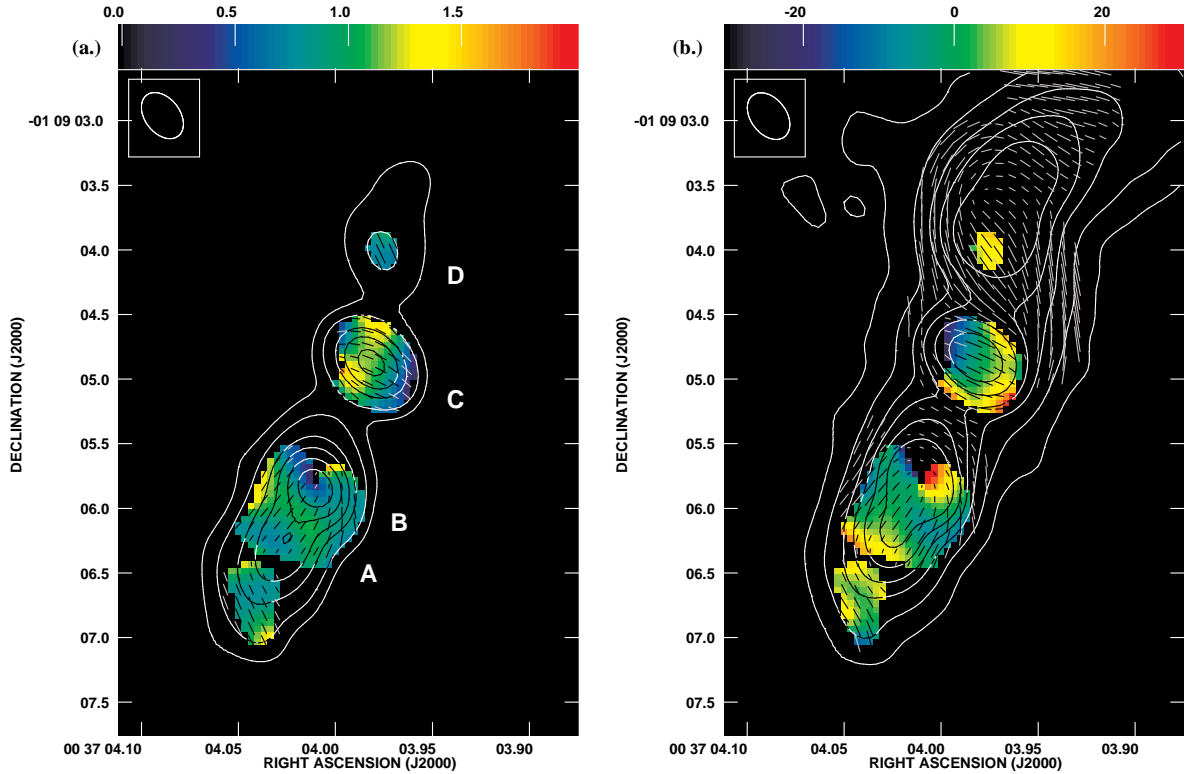}\\
\caption{Left (a.): Optical intensity contours and polarization vectors (rotated through 90$\degr$), after convolving the data of Figure~\ref{fig:data}b with the beam used to produce the radio images. Contour levels are $0.02~\times~(1, 2, 4, 6, 8.85, 12)$ electrons sec$^{-1}$, and a vector of length 0.2 arcsec corresponds to 50 per cent polarization. The colour scale shows the ratio between percentage of polarized flux detected in the radio and optical bands, where zero indicates relatively little polarized radio flux. Right (b.): Radio intensity contours and polarization vectors to the same angular scale as (a.). Contour levels are $0.2~\times~(1, 4, 10, 16, 24, 36)$ mJy beam$^{-1}$, and a vector of length 0.2 arcsec corresponds to 40 per cent polarization. The colour scale shows the difference between the optical and radio magnetic field position angles in degrees. See the electronic edition for a colour version of this figure.}
\label{fig:polarimetry}
\end{minipage}
\end{figure*}

The high-resolution Stokes $I$, $Q$ and $U$ {\it HST} images were first convolved to match the size and orientation of the radio beam before comparing the optical and radio data. The structure of the jet's apparent magnetic field is indicated in the optical and radio maps in Figure~\ref{fig:polarimetry}, where the polarization vectors have been rotated through 90$\degr$. Although this is common practice, these vectors do not necessarily show the true projected direction of the magnetic field, which may shift due to relativistic aberration \citep*[][and Section~\ref{subsec:model}]{Lyutikov03}. With this in mind, Figures~\ref{fig:data} and~\ref{fig:polarimetry} show that the radio and optical apparent magnetic field vectors in the central part of the jet tend to be perpendicular to the jet ridge line from knot~C outwards, but more parallel to the jet in the A/B region. Towards the edges of the jet the vectors tend to follow the flux contours. We note that Faraday rotation should be negligible at optical and 8.4~GHz radio frequencies.

High-resolution optical polarimetry (Figure~\ref{fig:data}b) shows that the apparent MFPA is initially at about $53\pm4\degr$ to the jet direction and that there are four distinct optical polarization minima in the knot~A/B complex. The first, and least well-defined minimum, is a relatively large region around knot~A spanning the whole width of the optical jet, after which the apparent MFPA rotates significantly to become parallel to the jet. The second minimum is more localised and occurs between knots~A and~B, while the third is located at the flux maximum of knot~B where the apparent magnetic field changes direction again to follow the flux contours. The overall pattern of the vectors in this region is reminiscent of a thumb-print. Downstream of knot~B, the apparent magnetic field becomes disordered and forms the fourth optical polarization minimum. At knot~C (also the brightest X-ray feature) the apparent MFPA vectors generally align perpendicular to the jet. Knot~D shows weak optical emission but exhibits similar polarization behaviour to that seen at the base of the optical jet, where the MFPA vectors are generally perpendicular to the new jet direction.

In the radio, the apparent MFPA at the base of the jet is broadly transverse, initially at about $60\pm2\degr$ to the jet direction. Just upstream of knot~A it rotates rapidly, becoming parallel to the jet, and aligns well with the optical MFPA. At knot~B itself the apparent MFPA changes rapidly around the polarization minimum, making the local fields difficult to interpret. Further down the jet, the radio MFPA is generally perpendicular to the jet direction along its centre-line.

\begin{figure*}
\begin{minipage}{160mm}
\centering
\includegraphics[width=120mm]{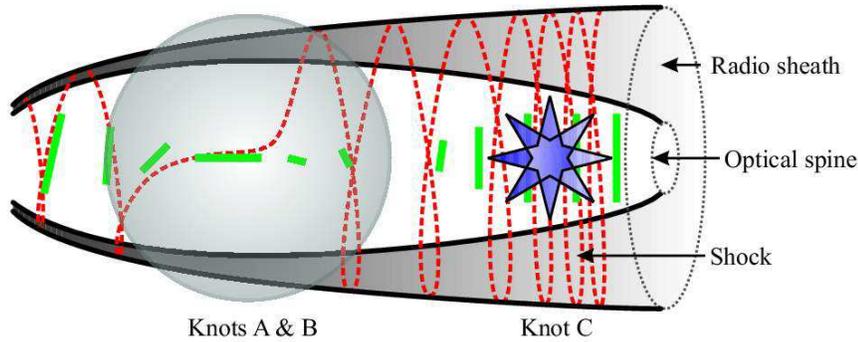}\\
\caption{A sketch of the inner jet, showing the main features out to knot~C. A possible helical magnetic field within the radio-emitting plasma sheath is shown by the dashed red line, although the details in the shaded region around knots~A and~B are not certain. The lines along the spine represent the apparent projected magnetic field vectors in the optical data.}
\label{fig:sketch}
\end{minipage}
\end{figure*}

Figure~\ref{fig:polarimetry}a (colour-scale) shows that the percentages of polarized emission seen in the optical and radio wavebands are similar throughout the jet, except in the region downstream of knot~B. Here the radio emission is significantly more polarized than in the optical, and the radio flux falls off less sharply with distance down the jet than the optical. At the base of the jet, the degrees of optical and radio polarization are almost the same ($25\pm3$ per cent in the optical and $21\pm3$ per cent in the radio) and they fall together to $5\pm1$ per cent at the first minimum just upstream of knot~A. The polarization ratio stays close to unity (both about $18\pm2$ per cent) between knots~A and~B, until the polarization direction changes at~B and the apparent magnetic field traced by the optical emission becomes disordered shortly afterwards. The radio data show a polarization minimum centred at knot~B. Knot~C shows a similar degree of polarization in both bands of $24\pm1$ per cent, although there is a small region to the east of the flux maximum where the optical polarization decreases sharply. Knot~D shows weak optical emission and is slightly more polarized in the optical ($40\pm6$ per cent) than in the radio ($33\pm4$ per cent).

The colour scale of Figure~\ref{fig:polarimetry}b shows differences in the apparent MFPA between the radio and optical data. An error map of the apparent MFPA differences was produced by measuring noise levels in both the radio and optical Stokes~$Q$ and~$U$ images, and including a $3\degr$ systematic uncertainty in the optical polarization position angle \citep[see the ACS Data Handbook,][for further details]{Pavlovsky05}. Regions where the statistical errors in the apparent MFPA are greater than $3\degr$ in the radio or $6\degr$ in the optical are blanked in Figure~\ref{fig:polarimetry} and are ignored in this analysis. It is clear that the apparent magnetic fields traced by both electron populations are quite similar, since the largest significant differences in field direction are only $10\degr$. In the knot~A/B region, where the signal-to-noise is good, the radio-optical MFPA differences are as large as $10\degr$, whereas the errors are $\sim4\degr$. Towards the edges of the jet the errors in the MFPA differences increase to $7\degr$. The downstream edge of knot~B is weakly polarized in the radio but unpolarized in the optical. Knot~C is different, as the apparent fields traced by the emission in both wavebands are generally perpendicular to the jet and are well-ordered. In the centre of knot~C the position angles align well ($0\pm4\degr$), but the offset is about $-10\pm5\degr$ at the eastern edge and $+10\pm5\degr$ to the west. The angles at knot~D are again about $10\degr$ apart, though we note that knot~D is detected with relatively low significance in the optical data.

\subsection{External gas}\label{subsec:external}
A radial profile of the X-ray image allows us to check for thermal emission from hot gas surrounding the core of 3C~15. Fitting a temperature to the spectrum and a beta-model to the profile provides an estimate of the thermal pressure at the position of knot~C. Care must be taken to exclude the non-thermal emission from the kpc-scale radio lobes in the extraction region, since these are also detected in X-rays \citep{Kataoka03} and are clearly visible in the 0.3--6~keV {\it Chandra} image. Two 125$\degr$ pie-slices were excluded, from position angles 270 to 35 degrees (covering the NW lobe), and from 90 to 215 degrees (the SE lobe). Annuli were defined so that each contained at least 20 counts, giving a total of 8 bins. There were 165 net counts out to a radius of 50 pixels (24.6 arcsec): background counts were obtained between radii of 50 and 80 pixels after applying the same exclusion criteria.

To probe for extended emission we need to test if the profile is consistent with the {\it Chandra} PSF. The PSF is energy-dependent, and was modelled using the X-ray spectrum of the core between 0.3--6~keV, using the {\sc chart} and {\sc marx} software from the CXC. The spectrum of the core is best fitted by a two-component power-law, using a circular extraction region of radius 1.3 arcsec. In agreement with \citet{Kataoka03} we find the photon index of the first component is $1.5 \pm 0.3$, while the second component is heavily absorbed ($N_H \sim 7 \times 10^{22}$ cm$^{-2}$) with slope $1.6 \pm 0.5$; $\chi^2 = 11.5$ for this fit, for $n = 21$ degrees of freedom. Formally the radial profile fits a model comprising only this PSF, with $\chi^2 = 8.6$ for $n = 7$. However, there is a trend towards positive residuals at the larger radii, suggesting the presence of a component of extended emission. A fit to a beta-model as well as a point-source component gives $\chi^2 = 2.4$ for $n = 4$, where $\beta = 1.5$ and the core radius $\theta_c = 10$ arcsec: both have large uncertainties. However, the formal significance of the extended component is low, with an F-test finding a 13 per cent probability of this improvement in $\chi^2$ from random data.

We have also used the spectral distribution of the counts in the lobe-free region to search for evidence of thermal emission. A fit using the spectral model of the core gives $\chi^2 = 3.45$ with $n = 5$ degrees of freedom ($P = 63$ per cent). Adding a {\sc raymond} thermal component gives a slight improvement of $\chi^2 = 2.02$ for $n = 4$ ($P = 73$ per cent; $kT \simeq 0.6$~keV, upper limit 0.9~keV). Checking the significance of the improved fit using the F-statistic shows a 17 per cent probability of obtaining the improvement by chance. Although this is not significant on its own, it complements our independent observation of weak extended emission seen at large radii and suggests that external gas may be present at a level close to our best-fitting value. Our beta-model and temperature fits are converted into pressure profiles using the equations of \citet{BW93}. We use the method of \citet{WB01} to find the uncertainty in pressure, but ultimately we treat our pressures as upper limits because of the weak evidence formally for the presence of the gas.

\section{Discussion}\label{sec:discussion}
\subsection{Magnetic structure?}
From Figure~\ref{fig:polarimetry} and the different widths of the optical and radio jets (Figure~\ref{fig:profile}), it is clear that the optical- and radio-emitting electrons are not entirely co-located. The data on the jet widths and the direction of the apparent magnetic-field vectors suggest the presence of an optically-bright spine and a lower-energy sheath. Spine-sheath structures in kiloparsec-scale jets have been proposed in many other sources: \citet{Laing96} summarises observational evidence for jets comprising a faster spine and a slower shear layer which decelerates due to the entrainment of external material. Jet velocity structures have also been proposed by \citet{Chiaberge00} in the unification of FR~I jets and BL~Lac objects, where a highly-relativistic spine is surrounded by a less-relativistic shear layer. These models require that the emission from the two components is only visible if it is beamed towards us. For a large number of sources, radio polarization measurements have shown that the apparent magnetic field is parallel to the jet near its base and that the field is generally perpendicular further out \citep{Bridle84}. Thus, in the region near the base of the jet, emission from the spine may be beamed out of our detectors, allowing us to see a parallel field from the shear layer. In 3C~15, the pattern of radio polarization vectors is consistent with what might be expected for a helical magnetic field in the outer component, and Figure~\ref{fig:sketch} shows a simple sketch of a possible magnetic field configuration in the inner jet based on the observed radio and optical data. We attempt to describe the 3C~15 jet using a more complex two-component model in a forthcoming paper.

Polarimetry around knots~A and~B shows very different behaviour to that seen at knot~C. The radio polarization minimum around knot~B may be an artefact of the vector geometry, where the field undergoes a rapid change of direction \citep*[a `lemon'; see][]{Scheuer77}, or the observed field may simply appear to be disordered along the line of sight and we are seeing the projection of small-scale field variations. It is clear that the field undergoes some rearrangement in knots~A and~B: both the radio and optical polarimetry (Figure~\ref{fig:data}) show similar abrupt changes in the apparent MFPA downstream of knot~A. The observations may be understood by a pattern shown in the shaded region of Figure~\ref{fig:sketch}, where a helical field in the sheath and in the spine has a reduced pitch angle, possibly because of a longitudinal expansion caused by a release of energy in the jet. However, it is also clear that there is almost no X-ray emission from this region. This may be because the shock strength, or field geometry, does not allow particle acceleration to X-ray emitting electron energies, even though bright X-ray emission (in the absence of optical and radio) is seen in A$'$. Another possibility is that the X-ray emission is sporadic, and has now ceased in knot~A.

The bright X-rays from knot~C indicate that it is a region of effective particle acceleration, possibly near a strong shock. Our radio and optical polarimetry support the shock model by showing a high degree of polarized emission from this region. The theoretical maximum polarization for synchrotron radiation is $\sim70$ per cent \citep{Westfold59,Pach70,RadioAstron} and the emissions from knot~C are about 25 per cent polarized. The apparent local magnetic field at~C is transverse to the jet and appears to be relatively well-ordered. One way to achieve this configuration is to enhance the local field by compressing plasma (containing frozen-in field lines) at a shock front. The question is then why is there a stronger shock at this location than at any of the other knots? An alternative view is that there is no shock and the apparent projected magnetic field just becomes well-ordered along the line of sight, although it is difficult to explain the relatively bright X-ray emission if this is all that is happening at knot~C.

\subsection{A test of a spine/sheath model}\label{subsec:model}

Our observations have constrained the differences in polarization position angle between the optically and radio bright regions of the jet. If these arise predominantly from regions of different velocity (as in a spine-sheath model), relativistic aberration of the polarization position angle may be large. In order to test such a model is not ruled out by the data, we attempt to model the jet as two components: a central high-velocity spine emitting optical synchrotron radiation, and an outer sheath of radio-emitting plasma. The simplest model has a magnetic field with components $(0, B'_y, B'_z)$ in the jet frame, where $y$ is transverse to the jet direction and $z$ is parallel to it, and the ratio of $B'_y$ and $B'_z$ is constant -- more complex models of the 3C~15 jet will be discussed in a later paper. Assuming a Lorentz factor $\gamma$ and an angle $\phi$ to the line-of-sight, $B_y = \gamma B'_y$ and $B_z = \big[\gamma + \sin^2\phi(1 - \gamma)\big] B'_z$ in the observer frame. One can then define $\tan\theta_{proj} = B_y/B_z$, where $\theta_{proj}$ is the projected angle of the magnetic field viewed by an observer relative to the jet, for $\phi>0$. This may be rewritten as
\begin{equation}
\tan\theta_{proj} = \frac{\gamma b'}{\gamma + \sin^2\phi(1-\gamma)}
\end{equation}
where $b'$ ($= B'_y/B'_z$) is the ratio of the two magnetic-field components in the jet frame. For a range of $\phi$, $b'$ and $\gamma$, we can examine the difference in $\theta_{proj}$ between an inner and outer component travelling at different speeds, and see if this can explain the difference in the polarization angle seen throughout the jet. We define $\theta_d$ as the difference in $\theta_{proj}$ between the two jet components. Values of $\theta_d$ are given in Table~\ref{tab:model} for a range of parameter values.

\citet{Leahy97} found a jet/counter-jet flux density ratio in 3C~15 of $\simeq50$, implying that $\beta\cos\phi = 0.635 \pm 0.025$ if the flux asymmetry is entirely due to a relativistic effect. This puts a constraint on $\phi \sim 45-50\degr$ for $\beta \ge 0.9$. If we assume the outer component is moving relatively slowly ($\gamma_o = 2$), the inner component is highly relativistic ($\gamma_i = 10$) and the jet lies at an angle of $50\degr$ to the line of sight, we can obtain $\theta_d=10\degr$ if $b'~=~1$. Figure~\ref{fig:polarimetry}b shows differences in the polarization angle between the optical and radio data of 0 to $\sim10\degr$ throughout much of the jet, so this is plausible. However, the errors in our measurements of $\theta_d$ are in the range $4-5\degr$ in the brightest regions of the jet, so it is difficult to constrain the parameters $b'$ and $\phi$ using these data. In any case, this simple magnetic field structure does not describe many of the polarization features we see, such as the rapid rotations in the MFPA seen separately in the optical and radio data. Thus, we cannot explain the 3C~15 jet purely in terms of a velocity difference between the two components and such a simple field configuration -- the field structure must itself vary, perhaps as in Figure~\ref{fig:sketch}.

If the jet contains a highly relativistic spine and a slower outer component, the optical jet/counter-jet flux density ratio would be different to that seen in the radio if the optically-emitting spine is moving much faster than the radio-emitting sheath. No counter-jet is detected in the optical image, and the optical jet/counter-jet ratio is not usefully constrained by the data: far deeper optical measurements are needed.

\begin{table}
\caption{Differences in the projected angle of the magnetic field ($\theta_d$) between the fast and slow jet components. Values are calculated for a range of viewing angles $\phi$ and field configurations $b'$, and assume the outer component is moving slowly ($\gamma_o = 2$) while the inner component is highly relativistic ($\gamma_i = 10$).}
\label{tab:model}
\begin{tabular}{lccccc}
\hline
$b'$ & $\phi=40\degr$ & $\phi=45\degr$ & $\phi=50\degr$ & $\phi=55\degr$ & $\phi=60\degr$ \\
\hline
0.25          & 4.2$\degr$  & 6.0$\degr$  & 8.4$\degr$   & 11.6$\degr$  & 15.8$\degr$ \\
0.5           & 6.3$\degr$  & 8.6$\degr$  & 11.4$\degr$  & 14.7$\degr$  & 18.3$\degr$ \\
1             & 6.3$\degr$  & 8.1$\degr$  & 10.0$\degr$  & 12.0$\degr$  & 14.0$\degr$ \\
2             & 4.2$\degr$  & 5.2$\degr$  & 6.2$\degr$   & 7.2$\degr$   & 8.1$\degr$  \\
4             & 2.3$\degr$  & 2.8$\degr$  & 3.3$\degr$   & 3.8$\degr$   & 4.2$\degr$  \\
\hline
\end{tabular}
\medskip
\end{table}

\subsection{Models for knot~C}\label{subsec:knot}
The X-ray spectral fits of Figure~\ref{fig:spectra} were obtained using calibration products from the Chandra X-ray Center which model the decline in the quantum efficiency of the {\it Chandra} ACIS detector at low energies. Consistent with this, our X-ray spectrum of knot~C is steeper than that reported in earlier work \citep{Kataoka03}, and we adopt the new value of $\Gamma_C=2.4 \pm 0.5$ for the photon index in this analysis. The result disfavours the SSC emission mechanisms that were previously suggested for knot~C, since SSC would have $\Gamma_C\sim1.7$, and we focus instead on modelling the spectrum using only synchrotron radiation.

\begin{figure*}
\begin{minipage}{160mm}
\centering
\includegraphics[width=70mm]{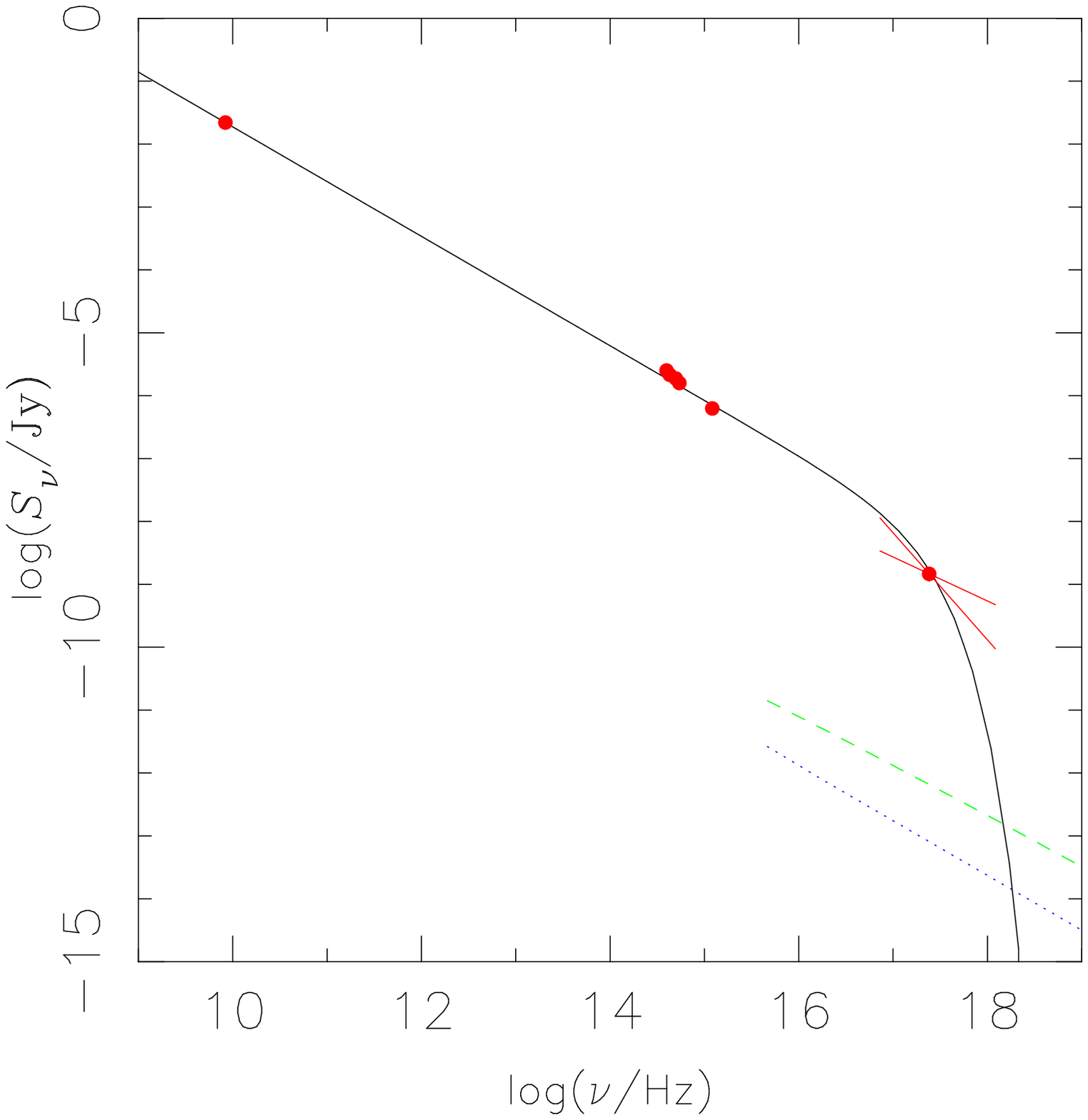}
\includegraphics[width=70mm]{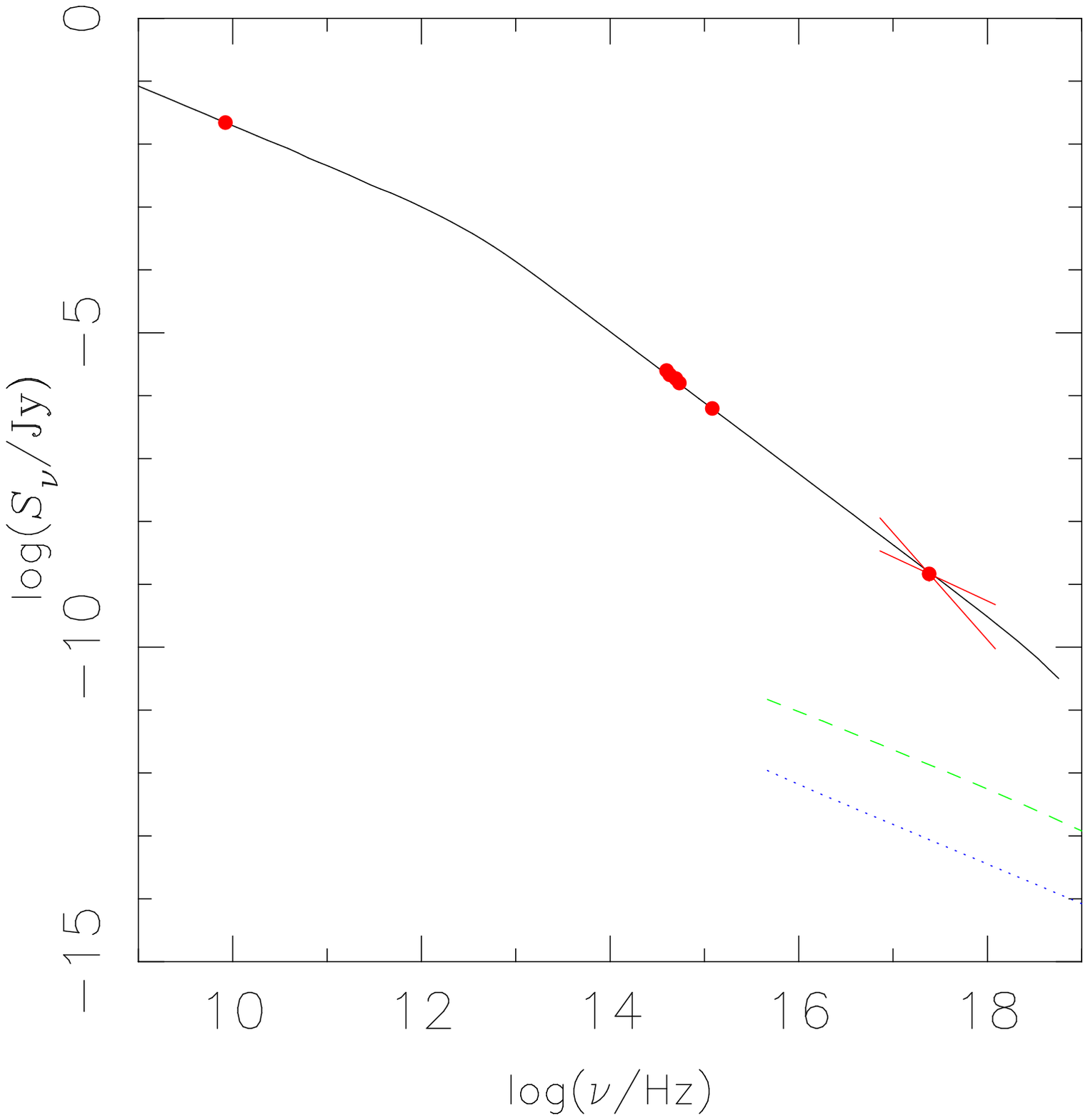}
\caption{Spectra for knot~C, comparing equipartition models of synchrotron radiation (solid line), synchrotron self-Compton emission (dashed line) and inverse Compton scattering of the microwave background radiation (dotted line). Left (a.): Simple power-law electron spectrum (slope $p=2.74$) with high-energy cut-off at $\gamma_{max}=1.2 \times 10^7$. Right (b.): Power-law electron spectrum with $p=2.26$, $\Delta p = 1.0$ at $\gamma_{brk} = 8.0 \times 10^4$.}
\label{fig:synch}
\end{minipage}
\end{figure*}

Radio, optical and X-ray flux densities at knot~C were derived directly from the images, and are listed in Table~\ref{tab:fluxes}. These data were used to fit synchrotron spectra to the knot emission. We find magnetic field strengths on the assumption that the radiating electrons and magnetic field are in equipartition, which is roughly equivalent to the source radiating with minimum total energy, and that relativistic beaming is negligible.

The derived spectrum (Figure~\ref{fig:synch}) shows the need to introduce a break in the synchrotron spectrum of knot~C, which is constrained by the X-ray photon index $\Gamma_C$. A model without a break, but with an exponential tail due to an upper cutoff in electron energy, would require an electron energy distribution with slope $p \sim 2.7$: this corresponds to a radio-optical spectral index $\alpha_{ro} \sim 0.85$ and implies a steep radio spectral index $\alpha_r$ of the same value. Such a model also requires a steeper X-ray photon index than we observe. The data effectively rule out this scenario, since the measured optical spectral index $\alpha_o = 1.25^{+0.01}_{-0.02}$ is much steeper than the predicted value of 0.85.

A model that includes a power-law break fits the data much better. We have constructed a synchrotron spectrum to fit the data using an electron power law of energy slope $p=2.26$, breaking by 1.0 at a Lorentz factor $\gamma_{brk} = 8.0 \times 10^4$; this is in line with models of acceleration at strong relativistic shocks, which predict $p=2.23$ \citep{Achterberg01} and $\Delta p = 1.0$ under a simple cooling assumption. This would also imply a radio spectral index $\alpha_r = 0.63$, consistent with values typical of most FR~I/FR~II jets, 0.5--0.7 \citep[e.g.][]{BP84}. Table~\ref{tab:synch} shows the parameters required for the fit shown in Figure~\ref{fig:synch}b, which is similar to that seen in other sources with kpc-scale jets \citep[e.g.][for 3C~66B and 3C~346 respectively]{Hardcastle01,Worrall05}.

\begin{table}
\caption{Synchrotron fitting parameters for the spectrum of knot~C, assuming equipartition. A fitted spectrum for each model is shown in Figure~\ref{fig:synch}.}
\label{tab:synch}
\begin{tabular}{lll}
\hline
Parameter & (a.) Power-law & (b.) Break of $\Delta\alpha = 0.5$ \\
\hline
$R$ (pc)                         & 250                   & 250                    \\
$\gamma_{min}$                   & 20                    & 20                     \\
$\gamma_{max}$                   & $1.2 \times 10^7$     & $1.2 \times 10^8$      \\
$p$                              & 2.74                  & 2.26                   \\
$\gamma_{brk}$                   & ...                   & $8.0 \times 10^4$      \\
$\nu_{brk}$ (Hz)                 & ...                   & $3.5 \times 10^{12}$   \\
$\Delta p$                       & ...                   & 1.00                   \\
$B_{eq}$ (nT)                    & 31                    & 20                     \\
$u_B$ (J m$^{-3}$)               & $3.7 \times 10^{-10}$ & $1.6 \times 10^{-10}$  \\
$u_{sync}$ (J m$^{-3}$)          & $1.1 \times 10^{-12}$ & $1.7 \times 10^{-12}$  \\
$t_{cool, 1 keV}$ (years)        & 16                    & 31                     \\
$t_{cool, \gamma_{brk}}$ (years) & ...                   & $8.5 \times 10^3$      \\
$P_{tot}$ (Pa)                   & $2.5 \times 10^{-10}$ & $1.1 \times 10^{-10}$  \\
\hline
\end{tabular}
\medskip
\end{table}

It should be noted that the radio and optical observations were taken almost 14 years apart. This should not change the position of any component by more than a few milliarcseconds, which is significantly smaller than the resolution of our data. However, finding the magnetic field in the jet allows us to estimate the synchrotron radiation loss-time, which is approximately $t_{cool} \sim 0.043B^{-3/2} \nu^{-1/2}$ years for an electron in a field $B$ Tesla emitting radiation at frequency $\nu$ Hz \citep[e.g.][]{Worrall06}. The equipartition magnetic field indicated in our best-fitting model is $B \simeq 20$ nT, which would give an electron radiative lifetime to 1 keV X-rays of $t_{cool} \simeq31$ years at knot~C; it is therefore clear that electrons must be reaccelerated in the knot and that the time between optical and X-ray observations is not important for our analysis. This field would also give a radiative lifetime of $t_{cool,\gamma_{brk}} \simeq8.5 \times 10^3$ years for electrons in knot~C emitting at the break frequency, which is comparable to the light-travel time from the nucleus (at least $1.7 \times 10^4$ years).

The total pressure in the plasma from fields and particles is found using $P_{tot} = 2 (\Gamma-1) B^2/2\mu_0$, where $\Gamma=4/3$ for a relativistic gas. Using the best-fitting equipartition magnetic field of 20~nT, we deduce the internal pressure of the knot region to be $\sim10^{-10}$ Pa. This may be compared with the upper limit on the thermal pressure $P_{ext} = 1.7 \times 10^{-12}$ Pa at the position of knot~C, based on the model for the external gas (Section~\ref{subsec:external}). If bulk relativistic motion is neglected, knot~C is overpressured by at least a factor $\sim60$.

The bulk Lorentz factor is $\delta = \sqrt{1- \beta^2} / (1 - \beta\cos\phi)$, where $\phi$ is the angle to the line of sight and $\beta=v/c$. Bulk relativistic motion will reduce the magnetic pressure in the knot by a factor $\delta^2$ \citep[e.g.][]{Worrall06}, and this may be estimated using the constraint $\beta\cos\phi = 0.635 \pm 0.025$ provided by \citet{Leahy97} based on the measured jet to counter-jet flux ratio on the assumption of twin jets that are intrinsically the same. Within the allowed ranges, $\delta$ can never be more than $\sim2.2$, which means the knot will still be overpressured by at least a factor of 10. It is possible that the two jets are intrinsically different and that the asymmetry is not simply due to beaming. Indeed, we already noted in Section~\ref{sec:intro} the evidence that the northern jet is FR~I-like while the southern is FR~II-like. However, taking the jet/counter-jet results at face value, our observations suggest the knot is likely to be a temporary feature in a state of expansion.

We note again that there is a small offset (0.14~kpc) between the positions of the intensity maxima at knot~C in the X-ray and radio images. Although this detection is barely significant (less than $2\sigma$) in the case of 3C~15, a similar offset has been observed in other nearby jets where the X-ray emission peaks closer to the nucleus \citep[e.g. $0.35 \pm 0.1$ arcsec or $0.21 \pm 0.06$ kpc in 3C~66B, as reported by][]{Hardcastle01}. If it is real, it may be the result of radio-emitting plasma being swept downstream, since the long lifetime of radio synchrotron emission would allow us to observe it at some distance from the shock front. Strong acceleration (to X-ray emitting energies) may only be permitted at the upstream edge of the shock. The other peculiar feature at knot~C is the small ($<0.1$ arcsec) offset between the X-ray and radio/optical flux maxima in the direction transverse to the jet. Unfortunately the resolution of {\it Chandra} does not allow us to claim that this offset is significant.

\section{Summary}
The bright, northern jet in 3C~15 shows similar characteristics to other FR~I class objects in showing a rich variety of different behaviours along its length. The base of the jet near knot~A$'$ is principally seen at high energies, in the X-ray and ultraviolet, before the optical and radio emissions turn on at knots~A and~B. Such behaviour is seen also in 3C~31 \citep{Hardcastle02} and NGC~315 \citep{Worrall06b}. With matched beams, the jet is also consistently narrower in the optical than it is in the radio for much of its length, suggesting that it is stratified. The exception is the A/B region, where the optical and radio jets are of similar width and brightness, although the region is not seen in X-rays. Throughout the jet, flux maxima appear to be accompanied by `pinches,' suggesting a new component of emission is appearing in the centre of the jet and making it seem narrower. Optical and radio polarimetry around~A and~B show polarization minima near flux maxima in both bands, also seen in parts of the M~87 jet \citep{Perlman99}. At knot~A there is a significant change in the magnetic field position angle, where the apparent field changes from being broadly transverse to the jet to being broadly parallel to it.

We have proposed a simple model of the synchrotron-emitting jet in 3C~15, consisting of an energetic, high-velocity spine and a radio-emitting sheath. However, the data do not allow us to constrain usefully the velocities of the flow. Using a simple magnetic structure we have checked that large velocity differences between a spine and a sheath would not produce a difference in radio and optical polarization position angles larger than we observe. The A/B region shows a greater variety of polarization structure than seen in the rest of the jet, suggesting that a particularly complex field configuration and/or velocity structure is present. Detailed magnetic modelling of the jet is deferred to a later paper. We note that 3C~15 has an unusual radio morphology, where the northern jet is dissipative and appears FR~I-like, while the counter-jet powers a `warm-spot' in the southern radio lobe and shows classic FR~II characteristics. For this reason we question the assumption that both jets are intrinsically the same, and that the observed flux asymmetry is entirely due to relativistic beaming: future models would then lose the constraint on $\beta\cos\phi$.

Knot~C is bright across the X-ray, optical and radio bands, and exhibits a high degree of optical and radio polarization where the magnetic-field position angle is relatively well-aligned in both components. We interpret these characteristics as the signatures of a strong shock with a well-ordered magnetic field, and high-energy emission indicates that it is a region of effective particle acceleration. Our synchrotron spectral model with equipartition magnetic field requires a break in the electron spectrum at $\gamma_{brk} = 8.0 \times 10^4$. The equipartition magnetic field of $10-20$~nT (depending on jet speed) implies knot~C is overpressured by $\sim15-60$ compared with the thermal pressure from external X-ray emitting gas. This suggests knot~C is a temporary, expanding feature. Knot~C is suitable for more detailed examination, since it is the only part of the jet showing emission in all wavebands.

Optical knot~D is detected for the first time in our long {\it HST} ACS/F606W+POL exposure, and shows a similarly high degree of polarization as knot~C. Knot~D is slightly more polarized in the optical than in the radio, although its optical emission is much weaker. Knot~D is not detected in our {\it Chandra} observation.

These results form part of an ongoing investigation into the magnetic environment of the jets of radio galaxies. 3C~15 is one of several sources with optical synchrotron jets (including 3C~66B, 3C~78, 3C~264, 3C~346 and 3C~371) where multi-band polarimetry is being assembled. Together with M~87, 3C~15 is one of the few kpc-scale jets with sufficient optical data that allows us to examine the magnetic structure in detail. Future multi-frequency radio observations and modelling will provide further insight into the structure of the jet, the nature of particle acceleration, and how these relate to magnetic field geometry.

\section*{Acknowledgments}
FD acknowledges a research studentship from the UK Particle Physics and Astronomy Research Council (PPARC), and thanks D.A. Evans, M.J. Hardcastle and J.H. Croston for helpful discussions. We also thank the anonymous referee for helpful suggestions to improve the manuscript. Work at UMBC was supported by NASA LTSA grant NNG05-GD63DG and {\it HST} guest observer grant GO-09847.01. The National Radio Astronomy Observatory is a facility of the National Science Foundation operated under cooperative agreement by Associated Universities, Inc. This research has used observations made with the NASA/ESA Hubble Space Telescope, obtained from the data archive at the Space Telescope Institute. STScI is operated by the association of Universities for Research in Astronomy, Inc. under the NASA contract NAS 5-26555. We thank the CXC for its support of {\it Chandra} observations, calibrations, and data processing.

\label{lastpage}

\end{document}